\documentclass[10pt,twocolumn]{article} 

\usepackage{simpleConference}
\usepackage{times}
\usepackage{graphicx}
\usepackage{amssymb}
\usepackage{url,hyperref}
\usepackage{comment}
\usepackage{float}
\usepackage[export]{adjustbox}
\usepackage{setspace}
 \usepackage{setspace}
\usepackage{caption}
\begin{document}

\title{Catalog of variable stars in the WD~0009+501  and GRW +708247 fields based on photometric survey data on transiting exoplanets}
\author{O.~Ya.~Yakovlev$^{*1}$, A.~F.~Valeev$^{1,2}$, G.~G.~Valyavin$^{1}$, V.~N.~Aitov$^{1}$, G.~Sh.~Mitiani$^{1}$, T.~A.~Fatkhullin$^{1}$,\\
G.~M.~Beskin$^{1,3}$, A.~V.~Tavrov$^{4}$, O.~I.~Korablev$^{4}$, G.~A.~Galazutdinov$^{2,1}$, V.~V.~Vlasyuk$^{1}$, E.~V.~Emelianov$^{1}$,\\
V.~V.~Sasyuk$^{3}$,  A.~V.~Perkov$^{5}$,  S.~F.~Bondar$^{\dag}$$^{5}$, T.~E.~Burlakova$^{1,2}$, S.~N.~Fabrika$^{1}$, I.~I.~Romanyuk$^{1}$
\\
\\
$^{1}$Special Astrophysical Observatory, Nizhnii Arkhyz, 
$^{2}$Crimean Astrophysical Observatory, Nauchny,\\
$^{3}$Kazan Federal University, Kazan, 
$^{4}$Space Research Institute, Moscow, \\
$^{5}$Research and Production Corporation ``Precision Systems and Instruments'', Moscow\\
$^{*}$yko-v@ya.ru\\
\\Accepted by Astrophysical Bulletin, December 7, 2023
}

\twocolumn[
\maketitle
\begin{abstract}
We present a catalog of 150 variable stars, including 13 stars with exoplanet candidates. 37 stars were identified as variables for the first time.  As a result of a 2.5-year photometric survey of exoplanets, we have obtained and analyzed light curves for almost 50 thousand stars in fields around white dwarfs WD~0009+501 and GRW~+708247.  Here we describe observations and data processing, the search for variable stars, their cross-identification with other catalogs and classification. The catalog is published in open access and contains the primary classification of variability, light curves and their parameters. 
\end{abstract}
]

\clearpage
\section{Introduction}
The detection and study of variable stars is an accompanying line of research in the process of searching for exoplanets by the transit method, the gist of which can be reduced to detecting variations in the light of stars. The chance of detecting new exoplanets increases with longer observations and larger ﬁelds considered. Light curves of tens of millions of stars are obtained as a result (e.g., the SuperWASP project \cite{Norton}), spanning several months or years. Catalogs are published and variable stars of diﬀerent types are studied based on the data of such space- and ground-based exoplanet surveys.

For example, over the course of the initial TESS space telescope mission in 2018–2020, a catalog was published (\cite{Fetherolf}) containing over 40 thousand variable stars. The light curves of about 200 thousand bright stars $m < 14^m $ were investigated in this survey on a time interval of approximately 56 days for a quarter of them and 28 days for the rest. TESS data were used to study variable A–F-type stars (\cite{Skarka}), short-period pulsating hot subdwarfs (\cite{Baran}), and eclipsing binaries (\cite{Green}). A catalog of 221 variable stars with m < 15m (\cite{Zhang}) was published based on the data of the Antarctic Survey Telescope (AST3-II) of the CHESPA project.

Photometric observations are conducted at SAO RAS as part of the EXPLANATION project (\cite{Valyavin_Photonics, 	Valyavin_AB}) using small robotic telescopes. Two of them are actively working at present, with another three mounted and being prepared for operation. In order to search for exoplanet candidates, observations of two regions with a size of about ${1.5^\circ}$ were carried out over the course of 2.5 years using one of these telescopes. As a result of these observations, photometric series with a maximum duration of 23 and 56 days were obtained for a total of almost 50 000 stars (for the ﬁrst and second region, correspondingly).

Considering the less than 10 per cent geometric detection probability of hot Jupiters (\cite{Deeg_2018}), the best attainable accuracy of 3  ${\sigma\approx0\,.\!\!^{\rm m}01}$, as well as the gaps between observations and their duration, we expect to discover based on these data no more than twenty new exoplanet candidates. At the same time, light curves for at least hundreds of variable stars have been obtained, including those previously unknown, that are suitable for study.

However, such an investigation of variable stars is not a priority for the EXPLANATION project. This work is therefore dedicated to the description of the compiled catalog of variable stars and their initial classiﬁcation, as well as the presentation of the light curve parameters so that these stars can be studied later by other research groups.

The catalog is published in open access at \href{https://www.sao.ru/jet/VarStarsDB}{link}\footnote{\url{https://www.sao.ru/jet/VarStarsDB/}} and will be updated in future. Currently it contains information about 150 eclipsing, pulsating and rotating periodic variables: the variability type, light curves of various degrees of processing, their parameters (period, amplitude) and phase folded light curves, as well as their data from other catalogs. 

\section{Observations and reduction}
The white dwarfs WD~0009+501 (EGGR-381) and GRW~+708247 (LAWD-73), studied earlier by the authors, were selected as central ﬁeld objects. For the observation latitude ($\phi\approx44^\circ$)  these stars are nonsetting ($\delta_1\approx50^\circ$, $\delta_2 \approx70^\circ$), which is important in an exoplanet survey for obtaining long photometric series.

The variability of EGGR-381 with a period of $P\approx8^h$  and amplitude $m\approx0\,.\!\!^{\rm m}01$  is reliably known (\cite{Valeev_2015, Antonyuk}). This white dwarf was chosen as a central object with the aim of testing the detection possibility for variations of such amplitudes, which are also typical for the maximum depth of hot Jupiter transits (\cite{Winn}). 

The variability of GRW~+708247 was studied by \cite{Valeev_2017}, however, it was not reliably conﬁrmed. The choice of this object as a central source is based on an assumption that it hosts a transiting exoplanet. 

Observations were carried out with the “Astrosib 500” robotic Ritchey-Chretien telescope with a $D=50$ cm main mirror. Frames were obtained in cumulative light in a  ($1\,.\!\!^{\prime\prime}3/$pixel) scale using a CCD detector with a 4096x4096 9-micron pixel array mounted in the central focus. The WD~0009+501 (GAIA mag. $G = 14\,.\!\!^{\rm m}2$) ﬁeld was observed for 84 nights from 2020/08/25 to 2021/01/21 with a 60 s exposure, and the GRW~+708247 ($G = 13\,.\!\!^{\rm m}2$) ﬁeld was observed for 224 nights from 2021/02/02 to 2022/12/31 with a 40 s exposure. 

The obtained frames were processed automatically by means of an algorithm of our design using standard photometric procedures: dark subtraction and (in the case of the second ﬁeld) ﬂatﬁelding (CCDPack, \cite{CCDPack}), locating the sources and their identiﬁcation in the frame with the reference catalog (Astrometry.net, \cite{Astrometry}), aperture photometry (\cite{SExtractor}), cross-identiﬁcation of sources and construction of instrumental light curves with their consequent calibration (CCDPack, \cite{astropy}). 

Gaia DR3 was adopted as a reference catalog for the coordinates (\cite{GAIAmission, GAIA3}); it was used to compile catalogs of almost 40 and 13 thousand stars from the observed ﬁelds in the $G\in[14\,.\!\!^{\rm m}5$--$19\,.\!\!^{\rm m}5]$ and e $G\in[11\,.\!\!^{\rm m}5$--$18\,.\!\!^{\rm m}5]$  intervals, correspondingly. Over the course of two surveys, photometric series for 35 183 and 12 589 stars with a length of up to 24 390 and 55 344 points were obtained that were later investigated for variations. 

Due to the fact that the quality of the photometric series was inhomogeneous, for various reasons, the methods described were used on the light curves in diﬀerent combinations with additional processing: varying the median level for each night, trend subtraction, removal of intervals of low accuracy due to bad weather conditions. The ﬁrst two procedures were applied to detect short-period variables or long-period eclipsing stars (i.e. sources with periodic variations that manifest at intervals shorter than $\sim6$ hours). 

Additionally, one to twenty reference stars were used for calibration in different instances, selected according to various criteria (see \cite{Yakovlev_2023}). Additional information about the observations and data reduction can be found in our earlier publications (\cite{Yakovlev_2022, Yakovlev_2023, Valyavin_AB, Valyavin_Photonics}). 

\section{Search for variable stars}

 To search for variable stars we used a step-by-step approach applied multiple times with various methods of studying the time series for variations and diﬀerent versions of light curve processing. A criterion was worked out at each step by which a star was either considered to be a candidate and studied further, or rejected.

The ﬁrst method of investigating for variability consists of constructing a dependence of the standard deviation of the light curve $\sigma$  on magnitude $m$ (\cite{Sokolovsky}). One might consider either individual nights, or the light curve as a whole. Those curves are selected that have deviations higher than expected for a given magnitude. They are then visually checked for the presence of variations or, in the case of a negative result, studied further. This method is suitable for searching for short-period or high-amplitude variables.

The main method used in the search consists of constructing and consequently analyzing the box least square (\cite{BLS}, \cite{astropy}) and Lomb–Scargle periodograms (\cite{Lomb}, \cite{Scargle}, \cite{LS}, \cite{astropy}). The ﬁrst algorithm is more effective in the search for Algol-type eclipsing variables and exoplanet transits, whereas the second method is best for other stars (with smoothly changing light curves). Among the obtained periodograms we select those that have maximum signal-to-noise ratios or maximum signal-to-other peak ratios higher than some limiting value. Phase light curves are then constructed for the selected objects for periods with the highest mathematical signiﬁcance and their multiples, after which they are visually analyzed.

A small number of variable stars were discovered by accident during the process of tuning the light curve creation algorithm and their calibration, as well as in the process of selecting the reference stars. Periodograms were then also plotted for them with the aim of determining the period and constructing the phase light curve with subsequent testing. 

\section{Identification with other catalogs}
 After the used algorithms had stopped yielding new results, the stars in the studied ﬁelds were identiﬁed with the known variable stars (or candidate variables) from other published catalogs. In total, we considered eight variable star catalogs presented in Table~\ref{table:table1}, as well as the exoplanet catalog\footnote{{NASA Exoplanet Archive}:\\ \url{https://exoplanetarchive.ipac.caltech.edu/}} . Two of these (the last in Table~\ref{table:table1} ) have to records for the given coordinates and magnitude range, as is the case with the exoplanet catalog.

\begin{table*}[h!]
	\centering
	\caption \centering{Used catalogs of variable stars. Shown are the number of stars in the catalogs\\in the ﬁelds considered and the number of those conﬁrmed based on the investigated data}
	\begin{tabular}{c|c|c|c}
		\hline
		№ & Name	& Amount &Link to the Table in Vizier\\	
		\hline
		1 & GAIA DR3 Variables	&	911/101	&I/358/vclassre \cite{GAIA3} \\
		2 & VSX (AAVSO)	&	164/68 & B/vsx/vsx \cite{Watson} \\
		3 & ZTF			&	128/60 &J/ApJS/249/18/table3 \cite{Chen} \\
		4 & ASAS-SN		&	16/47	&II/366/catalog \cite{Jayasinghe} \\
		5 & ZTF	(suspected)	&	218/2 &J/ApJS/249/18/table2 \cite{Chen} \\
		6 & GCVS		&	3/2	&B/gcvs/gcvs\_cat \cite{Samus}	 \\
		7 & OGLE-V		&	0/0	&J/AcA/62/219/list\_var \cite{Soszy} \\
		8 & Zhang		&	0/0	&J/ApJS/240/16/table4 \cite{Zhang}\\
		\hline
	\end{tabular}
	\label{table:table1}
\end{table*}

A total of 1068 conﬁrmed or candidate variables are included in the selected catalogs in the ﬁelds considered, of which 717 are located in the ﬁrst ﬁeld, and 351 in the second. Most (911) are candidates in the Gaia DR3 variable star catalog (No. 1 in Table~\ref{table:table1} ). Variability has been conﬁrmed for only $n_{cat}=113$ of them according to our data (87 in the WD~0009+501 ﬁeld and 26 near GRW~+708247).

\begin{figure*}[]
	\begin{center}
	\includegraphics[scale=0.275]{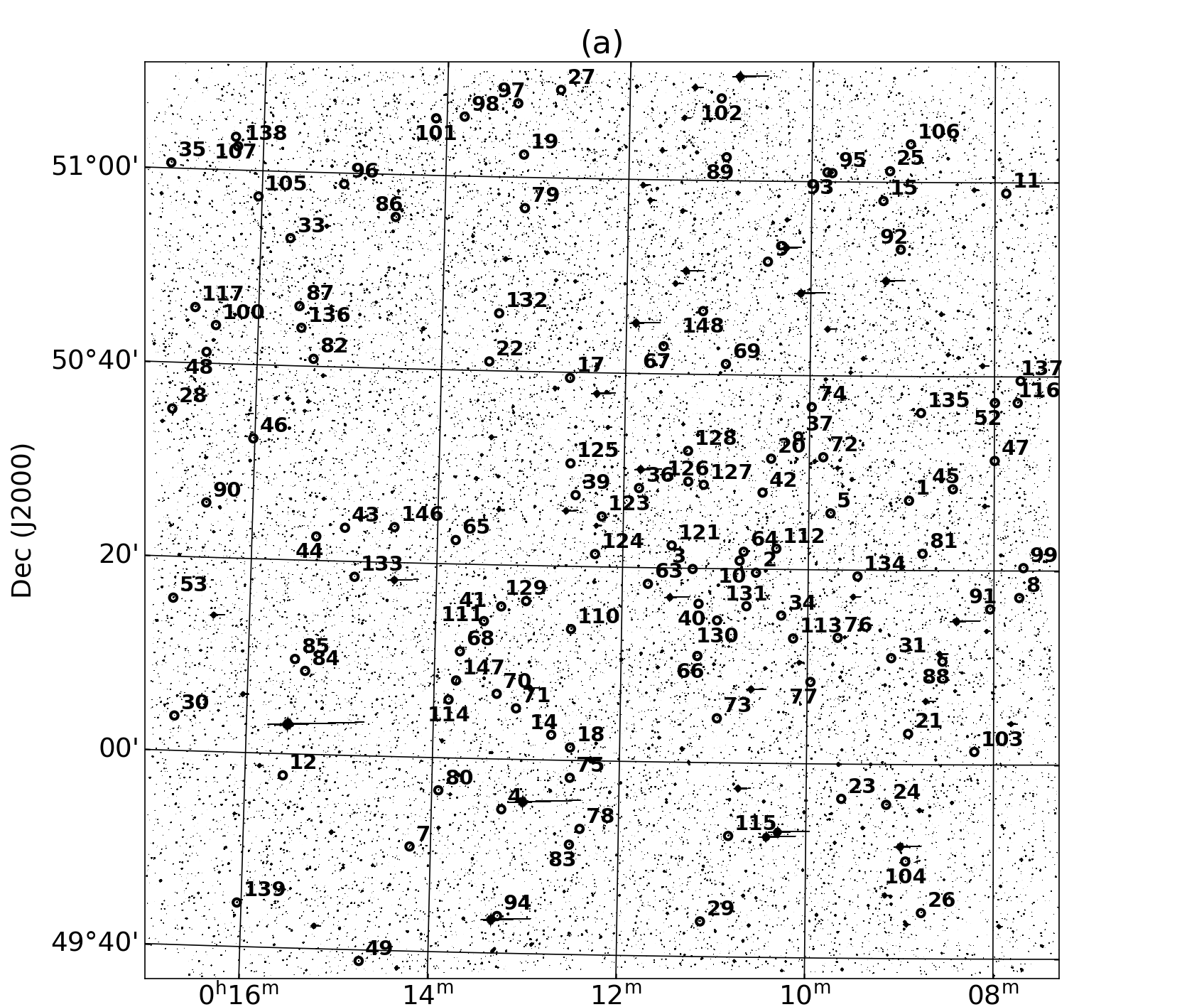}
	\includegraphics[scale=0.275]{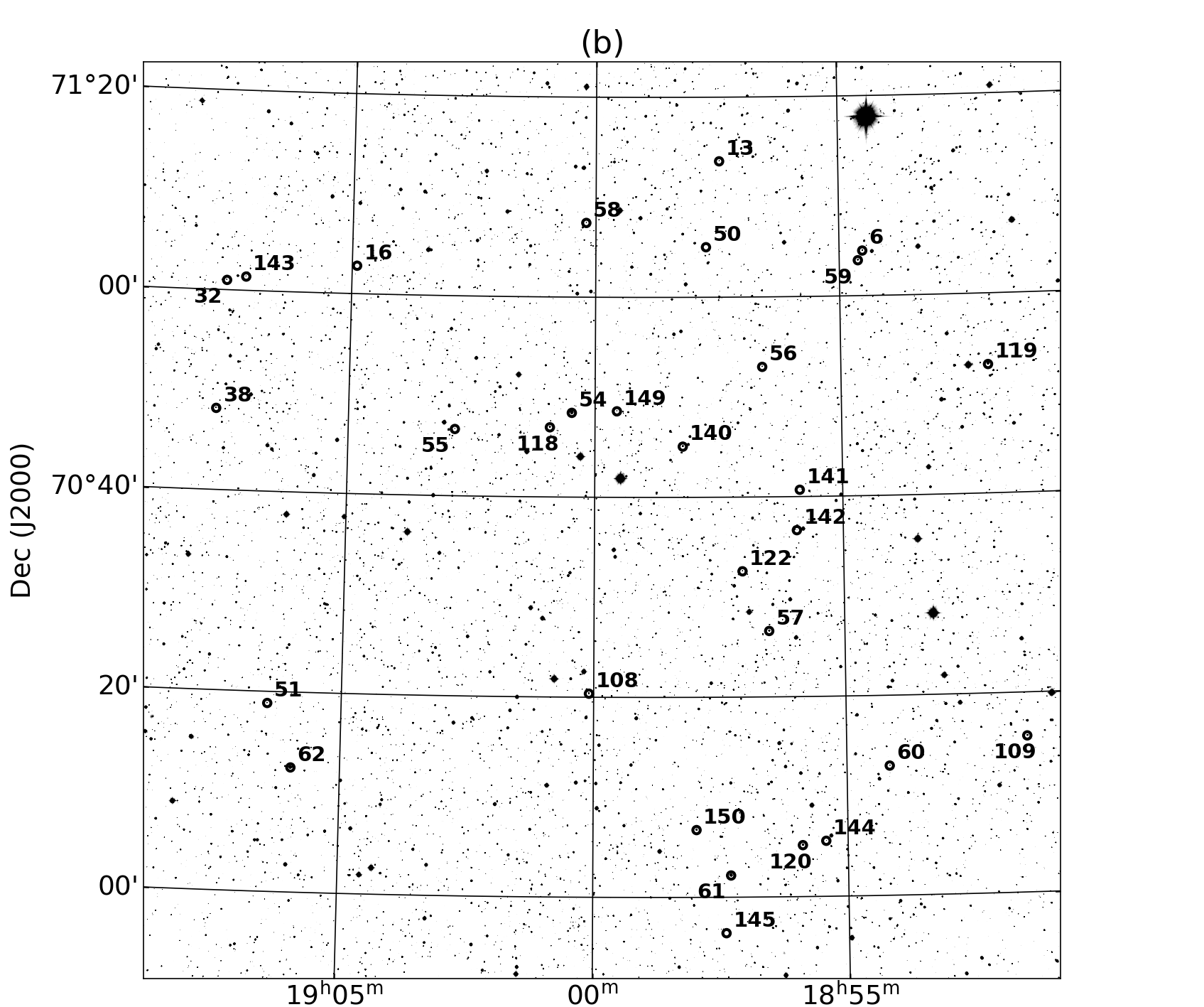} 
	\end{center}
	\caption{Positions of the variable stars from the catalog in the WD~0009+501 ﬁeld (a) and GRW~+708247 (b)  ﬁelds.}
	\label{fig:figure1}
\end{figure*}

Using the methods described above (see Section 3) a total of 124 variable stars have been found, $n_s=86$  of which were already included in other catalogs, and the remaining 37 were not. To these 123 stars we added the remaining $n_{cat}-n_s=27$ stars and obtained 150 variables in total for our catalog (Fig.~\ref{fig:figure1}).

The search for variable stars did not have a systemic nature, and we therefore give a rough estimate for the eﬃciency of the used approach as a ratio of $n_s/n_{cat}= 76\%$. In this estimate we do not take into account stars that were unknown previously or discovered by accident, and also the false negative variables from other catalogs (falsely not conﬁrmed by the investigated data). 

\section{Classification}
\begin{figure*}[h!]
	\begin{center}
		\includegraphics[scale=0.42]{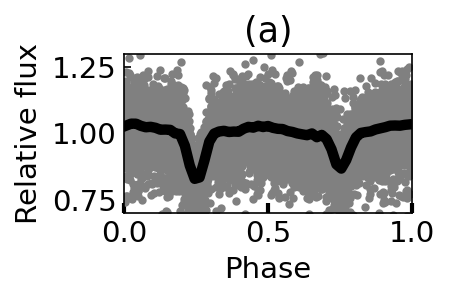} 
		\includegraphics[scale=0.42]{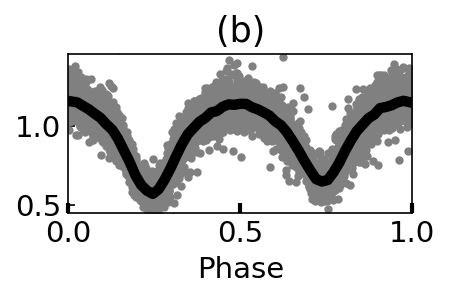} 
		\includegraphics[scale=0.42]{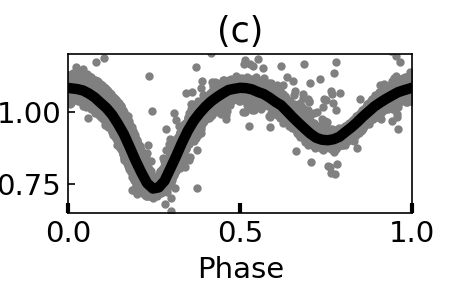} 
		\includegraphics[scale=0.42]{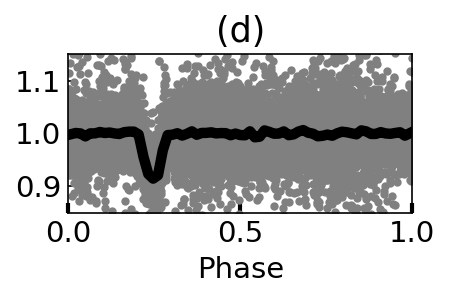}\\
		
		\includegraphics[scale=0.42]{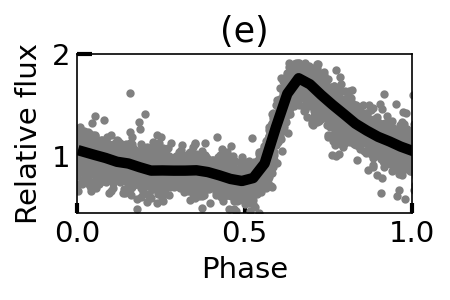} 
		\includegraphics[scale=0.42]{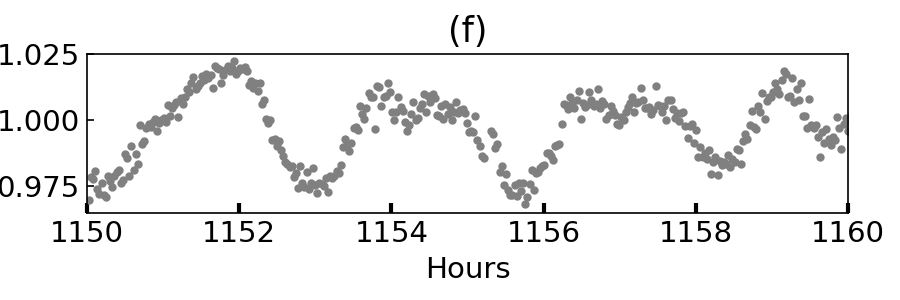}
		\includegraphics[scale=0.42]{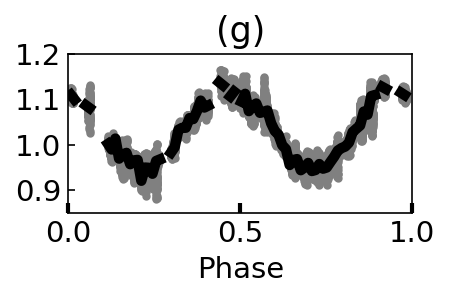}\\
		
		\includegraphics[scale=0.3]{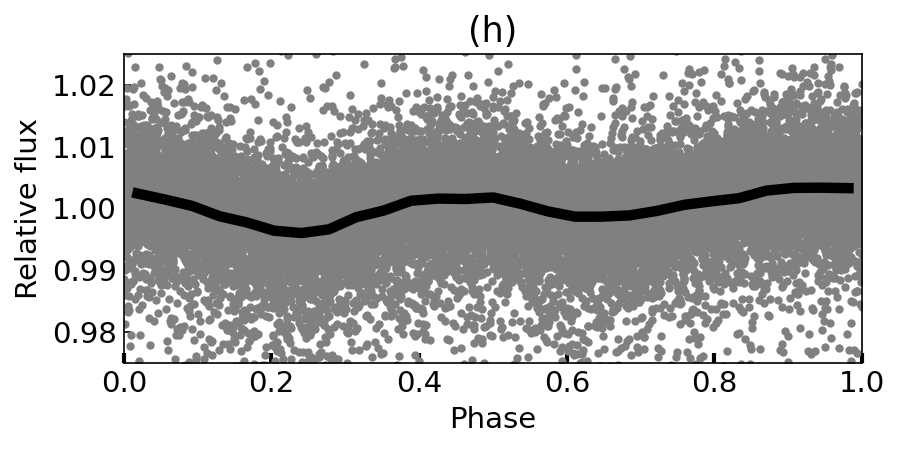} 
		
	\end{center}
	\caption{Phase folded light curve examples for variable stars: (a)~Algol~(EA), (b)~$\beta$~Lyrae~(EB), or (c)~W~Ursae Majoris~(EW) type eclipsing variables; (d)~stars with exoplanet candidates (EP$\vert$BD$\vert$EA); variables of the (e)~RR~Lyrae~(RR), (f)~$\delta$~Scuti or $\gamma$~Doradus type, or SX~Phoenicis type (DSCT$\vert$GDOR$\vert$SXPHE), (g)~long-period variables (LPV); (h)~rotating variables (rot).
	}
	\label{fig:figure2}
\end{figure*}

The classiﬁcation was carried out according to the “General Catalog of Variable Stars” (GCVS) (Samus’ et al., 2017) and “The International Variable Star Index” (VSX2). The variables found in the ﬁelds considered are divided into three main classes (Fig.~\ref{fig:figure2}): eclipsing (a)–(d), pulsating (e)–(g), rotating (h). The eclipsing and pulsating classes are subdivided into 4 and 3 types, correspondingly.

\begin{figure*}[ht!]
	\includegraphics[scale=0.6]{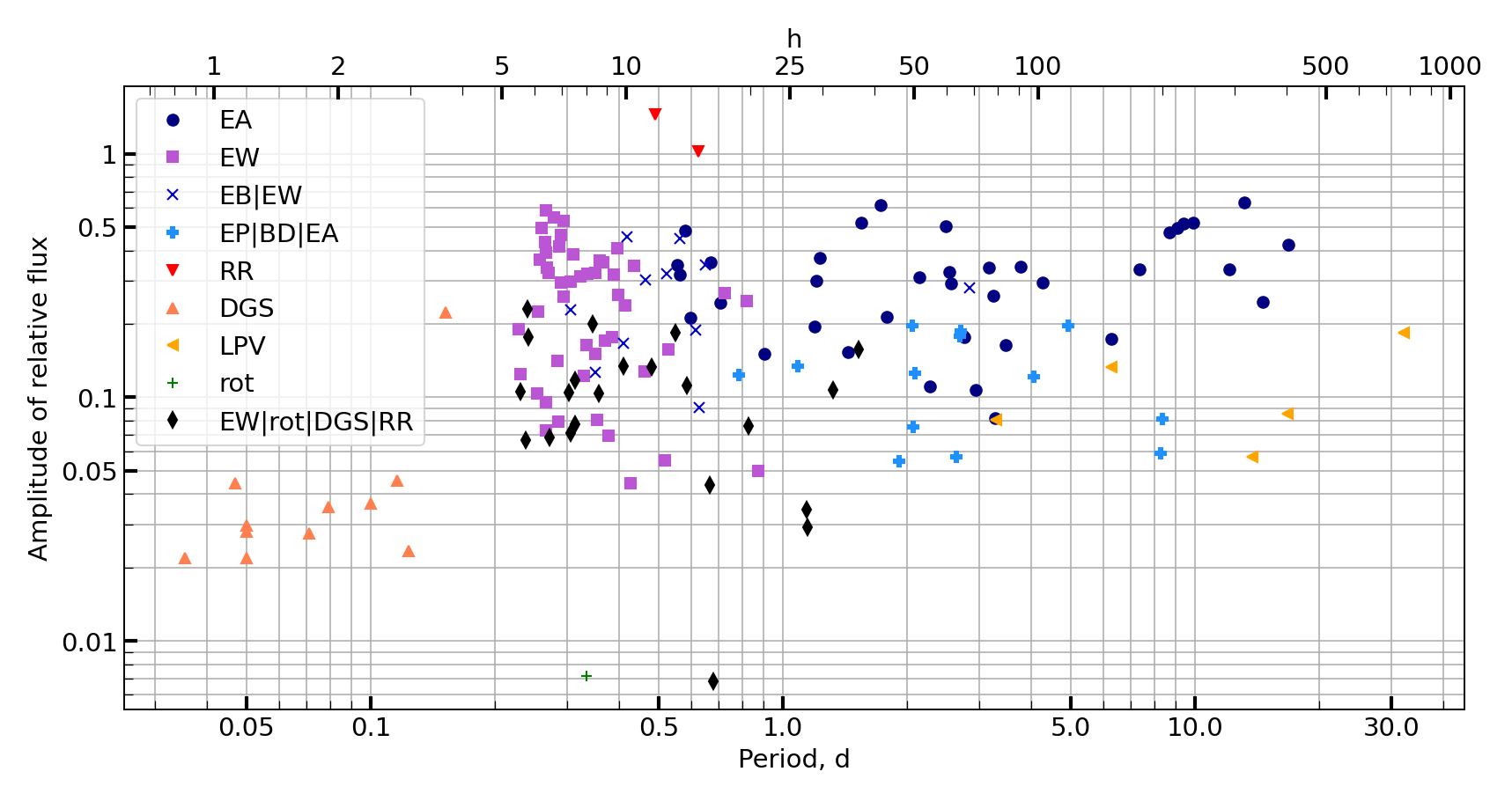}
	\caption{Variable stars of the catalog in the “period–amplitude” plane classiﬁed by type: eclipsing EA, EW, EB$\vert$EW, EP$\vert$BD$\vert$EA; pulsating RR, DSCT$\vert$GDOR$\vert$SXPHE (designated DGS), LPV; rotating rot, as well as a group of unambiguously unidentiﬁed types EW$\vert$rot$\vert$DGS$\vert$RR (see text for description).}
	\label{fig:figure3}
\end{figure*}

Three generally accepted types are isolated among the eclipsing variables (E): Algol (EA), $\beta$~Lyrae (EB), W Ursae Majoris (EW); and, separately, stars with exoplanet candidates (EP$\vert$BD$\vert$EA). If the presence of a transiting exoplanet is conﬁrmed for a star, it can be classiﬁed as an EP type object. Otherwise, if the eclipsing component is a brown dwarf, it belongs to the BD type (isolated only in the VSX system) or the EA type, if the second component is a star.

Three groups are isolated in the pulsating variables class: RR Lyrae (RR); $\delta$~Scuti or $\gamma$~Doradus, or SX Phoenicis (DSCT$\vert$GDOR$\vert$SXPHE); long-period variables (LPV) with periods longer than three days.

The VSX, ASAS-SN and ZTF catalogs also use such a classiﬁcation by type (with the exception of the fourth type of eclipsing stars). In Gaia the eclipsing type (E) is not subdivided into types; the RR Lyrae (RR) and solar (SL) types are present, and also there is no distinction between the $\delta$~Scuti, $\gamma$~Doradus and SX Phoenicis (DSCT$\vert$GDOR$\vert$SXPHE) types. Additionally, all catalogs isolate the RS Canum Venaticorum (RS) type, which is identiﬁed with eight stars with the numbers 15, 95, 102, 104, 129, 133, 146, 149. The light curves that we obtained do not conﬁrm their belonging to this type (see the light curves in Supplementary Materials).
\\
\\
Out of the 113 variables contained in other catalogs, 111 were already classiﬁed, and the remaining two from catalog No. 5 (Table~\ref{table:table1} ) were not. Based on this information and the obtained observational data (shape, period and amplitude of the phase curve) a decision was made with regard to which type the star belongs to. In the case of a discrepancy between the catalogs, VSX, ASAS-SN and ZTF were preferred due to the fact that the Gaia classiﬁcation (unlike these catalogs) was conducted fully automatically. Using an automatic classiﬁer (Kim and Bailer-Jones, 2016), implemented using machine learning, did not yield signiﬁcant results. Periods $P$ and variation amplitudes $\Delta m$ were determined for all catalog stars; they are presented in Table~\ref{table:table2}  and in Fig.~\ref{fig:figure3}.

\section*{Conclusion}
A catalog of 150 stars has been compiled as a result of a study of light curves of the stars in the WD~0009+501 and GRW~+708247 ﬁelds using variation search algorithms and variable star catalogs. It contains 109 eclipsing variables (type E), most of which (47) are related to type EW, 38 to type EA, 11 to EB$\vert$EW and 13 to stars with exoplanet  candidates EP$\vert$BD$\vert$EA. The pulsating star type (Puls) applies to 18 variables, of which 2 are RR type, 11 are (DSCT$\vert$GDOR$\vert$SXPHE) and 5 are LPV. One star (WD~0009+501) is classiﬁed as rotating (rot), and the remaining 22 belong to the inconclusively deﬁned EW$\vert$rot$\vert$DGS$\vert$RR type. 
\\
\\
We thank the anonymous reviewers  for their useful comments that helped to improve the manuscript and L.~Chmyreva for translation it from Russian. The analysis of variable objects was carried out with the support of the Russian Science Foundation grant 23-62-10013.

\begin{table*}[!ht]
\caption{Parameters of the stars from the catalog presented in this work and their variations. In the last column the type is given for stars known from other catalogs. DGS is the abbreviation used for (DSCT$\vert$GDOR$\vert$SXPHE). Online and .csv versions are available \href{https://www.sao.ru/jet/VarStarsDB}{here}. Eight exoplanet candidates published in \cite{Yakovlev_2023} were noted as SOI-n. 
}
	\label{table:table2}

	\begin{tabular}{c|c|c|c|c|c|c|c|c|c}
		\hline
		index&GAIA DR3 source id&RA&Dec&mag G&Amp&Per&Type&Type-0\\
		\hline
		\multicolumn{8}{c}{\textbf{Class: Eclipsing}}\\
1&395274812445302016&2.229875&50.453245&15.5&0.24738&351.289&EA&\\
2&395223478994996608&2.640593&50.327162&14.9&0.49551&218.0&EA&E\\
3&395221314331486848&2.811233&50.332251&14.19&0.17954&65.56&EA&E\\
4&395193856599122560&3.313128&49.913861&17.49&0.5043&59.8138&EA&E\\
5&395272579062402816&2.441462&50.430547&14.66&0.1733&150.2878&EA&E\\
6&2265877140230153600&283.630864&71.075051&15.59&0.29491&102.68&EA&EA\\
7&394444818601580928&3.555535&49.846449&15.73&0.26223&78.0296&EA&E\\
8&393778308397569920&1.931906&50.286905&14.45&0.52032&37.2892&EA&EA\\
9&395355008075442688&2.617854&50.861803&17.38&0.21355&42.9687&EA&E\\
10&395224264969539584&2.685054&50.347333&18.36&0.3398&76.1074&EA&\\
11&395324324828879488&1.969409&50.98153&17.42&0.29458&61.5569&EA&\\
12&394459077895183104&3.898685&49.962845&19.02&0.32565&60.902&EA&\\
13&2262905160300329728&284.364162&71.226466&16.35&0.19505&28.62&EA&E\\
14&395201621900114176&3.183671&50.043319&17.56&0.21223&14.3&EA&EA\\
15&395312779956817280&2.304452&50.967816&14.58&0.10711&70.5221&EA&RS\\
16&2262840868933718528&286.225297&71.046513&18.14&0.47657&208.4&EA&\\
17&395336625615598720&3.1509&50.657034&14.97&0.17602&66.0876&EA&\\
18&395204095801255040&3.132701&50.022202&14.5&0.37378&29.5164&EA&EA\\
19&395370160720961920&3.287777&51.039151&16.3&0.33487&176.34&EA&\\
20&395249798555153408&2.603089&50.523399&17.05&0.33628&290.8016&EA&\\
21&393761880145088000&2.228665&50.05279&18.34&0.62966&317.0946&EA&\\
22&395334594088690304&3.370551&50.682601&18.55&0.42489&405.6&EA&\\
23&393752328136796800&2.406077&49.940379&18.64&0.30118&28.999&EA&\\
24&393755042556557056&2.286177&49.930757&13.63&0.30994&51.4729&EA&EA\\
25&395406685123553664&2.286832&51.019033&15.12&0.08181&78.4668&EA&\\
26&393727387266077696&2.19181&49.745241&15.34&0.11052&54.6465&EA&\\
27&395374043371349248&3.189686&51.15151&18.73&0.3199&13.4886&EA&E\\
28&394593527545967360&4.226079&50.586942&17.54&0.15319&34.6235&EA&\\
29&393699452798541824&2.778971&49.7272&16.05&0.34961&13.3426&EA&E\\
30&394463579019096064&4.19349&50.059788&17.37&0.48545&13.885&EA&E\\
31&393764358343856512&2.275185&50.182619&15.61&0.15133&21.7147&EA&E\\
32&2262828709882555264&286.889657&71.016243&16.11&0.24376&16.905&EA&E\\
33&394607271444400384&3.919518&50.885796&15.44&0.52097&238.1&EA&\\
34&395217118142344064&2.571486&50.254179&17.94&0.3451&90.9012&EA&E\\
35&394982067472137984&4.250933&51.009383&13.71&0.16321&83.475&EA&E\\
36&395246534380053760&2.95904&50.469869&16.83&0.35978&16.0243&EA&E\\
37&395297008837063936&2.530646&50.562157&17.2&0.61332&41.4194&EA&\\
38&2262822525129532160&286.921449&70.802437&15.47&0.51731&225.05&EA&EA\\
39 (SOI-2)&395239868590825472&3.129905&50.455905&18.87&0.12539&50.358&EP$\vert$BD$\vert$EA&\\
40 (SOI-4)&395220833295162112&2.794095&50.272619&17.84&0.05672&63.422&EP$\vert$BD$\vert$EA&\\
41 (SOI-5)&395229079635160320&3.324483&50.262183&14.92&0.05902&198.309&EP$\vert$BD$\vert$EA&\\
42 (SOI-1)&395249244499720320&2.625306&50.464983&18.82&0.13406&26.085&EP$\vert$BD$\vert$EA&\\
43&394576102863768832&3.749979&50.390906&18.71&0.1788&64.5&EP$\vert$BD$\vert$EA&E\\
44&394564111315029120&3.826784&50.374682&16.69&0.08168&200.1872&EP$\vert$BD$\vert$EA&\\
45 (SOI-3)&395281753112434304&2.111531&50.473281&14.26&0.05474&45.988&EP$\vert$BD$\vert$EA&\\
46 (SOI-8)&394573186584164096&4.004353&50.540019&16.97&0.12397&18.768&EP$\vert$BD$\vert$EA&\\
47 (SOI-6)&395282813962513536&1.998838&50.522187&17.98&0.07576&49.645&EP$\vert$BD$\vert$EA&\\
48 (SOI-7)&394596585565904000&4.138275&50.686119&18.26&0.12097&97.67&EP$\vert$BD$\vert$EA&\\
49&394436915861885568&3.683293&49.647688&15.33&0.1968&49.45&EP$\vert$BD$\vert$EA&E\\
50&2262876259464278656&284.434002&71.083422&18.58&0.1968&118.1&EP$\vert$BD$\vert$EA&\\
	\hline
\end{tabular}
\end{table*}

\begin{table*}
\begin{tabular}{c|c|c|c|c|c|c|c|c|c}
\hline
№&GAIA DR3 source id&RA&Dec&mag G&Amp&Per (days)&Type&Type-0\\	
\hline
51&2262779919053939584&286.615606&70.313854&16.02&0.1865&64.85&EP$\vert$BD$\vert$EA&\\
52&395289415334015360&1.998741&50.621772&14.57&0.18949&14.74&EB$\vert$EW&EW\\
53&394560336042255744&4.206689&50.262103&14.12&0.34982&15.555&EB$\vert$EW&EW$\vert$EB\\
54&2262857950020261120&285.117959&70.807093&11.48&0.09064&15.0836&EB$\vert$EW&EB\\
55&2262810705379457664&285.709715&70.777887&17.04&0.1268&8.414&EB$\vert$EW&EW\\
56&2262870147726923776&284.150449&70.883421&16.28&0.22868&7.341&EB$\vert$EW&EW\\
57&2262479546221314176&284.127211&70.44319&14.09&0.28055&68.03&EB$\vert$EW&EA\\
58&2262888972567506560&285.050036&71.123747&16.67&0.4579&10.0512&EB$\vert$EW&EW\\
59&2265875456603234176&283.65437&71.059045&14.78&0.16653&9.846&EB$\vert$EW&EB$\vert$EW\\
60&2262449309651289984&283.539658&70.216066&16.37&0.4483&13.496&EB$\vert$EW&EW\\
61&2262432469083531264&284.324296&70.036215&11.85&0.30426&11.1316&EB$\vert$EW&EW\\
62&2262403130163052928&286.4915&70.207661&11.81&0.32283&12.5229&EB$\vert$EW&EW\\
63&395243716881546496&2.931151&50.305563&17.04&0.07926&6.8423&EW&EW\\
64&395224303628713216&2.674422&50.362957&14.26&0.38475&7.443&EW&EW\\
65&394486183430499072&3.450808&50.374492&17.87&0.12285&7.9182&EW&EW\\
66&395218462473241088&2.795403&50.182924&16.29&0.26878&17.3508&EW&EW\\
67&395255704130776960&2.898425&50.714223&14.95&0.41457&6.882&EW&EW\\
68&394477391633388032&3.432817&50.183615&16.81&0.31247&7.7352&EW&EW\\
69&395257632575459456&2.728605&50.685036&16.09&0.23825&9.9781&EW&EW\\
70&395226120393665920&3.33176&50.111913&18.04&0.29692&6.9625&EW&EW\\
71&395202760072882176&3.279247&50.087632&18.95&0.34094&6.421&EW&EW\\
72&395296768318905600&2.462304&50.526814&16.84&0.17667&9.2266&EW&EW\\
73&395210628452927872&2.741097&50.076362&16.06&0.12788&11.1212&EW&EW\\
74&395298898622664064&2.494746&50.612826&17.42&0.36794&6.168&EW&E\\
75&395198048487250688&3.131944&49.970174&17.34&0.12468&5.5476&EW&EW\\
76&395263022752784128&2.419259&50.216812&16.66&0.39183&6.404&EW&EW\\
77&395261442204729728&2.491278&50.140583&17.52&0.4941&6.2454&EW&EW\\
78&395195918183399808&3.103369&49.882752&18.31&0.4639&6.967&EW&EW\\
79&395367961697722112&3.282224&50.947198&17.68&0.31962&9.3532&EW&EW\\
80&394450144360969856&3.48227&49.944036&15.69&0.35738&8.8098&EW&EW\\
81&395269040009279872&2.192487&50.36224&17.38&0.5288&7.072&EW&E\\
82&394598578430693888&3.847475&50.679567&16.49&0.22431&6.112&EW&EW\\
83&395195819405788672&3.130853&49.855421&14.31&0.05513&12.4284&EW&E\\
84&394467049352650112&3.846734&50.143194&17.26&0.43445&6.3745&EW&EW\\
85&394468526820932736&3.874893&50.163183&18.39&0.41006&9.5022&EW&EW\\
86&394617471990323200&3.634264&50.926918&17.72&0.07293&6.4&EW&\\
87&394600743094163712&3.889721&50.76947&17.52&0.29855&7.338&EW&\\
88&393765075598535296&2.137554&50.177131&17.17&0.32087&8.0628&EW&E\\
89&395362842095742848&2.733725&51.04038&17.45&0.17051&8.8774&EW&EW\\
90&394568170064248704&4.126014&50.427404&14.13&0.06953&9.0472&EW&n/a\\
91&393777380684644480&2.010245&50.267208&15.85&0.24783&19.6766&EW&EW\\
92&395310546574009344&2.256365&50.884473&16.91&0.26386&9.5594&EW&E\\
93&395403798905544576&2.457903&51.016252&17.16&0.32463&6.4868&EW&EW\\
94&395190905963218944&3.318667&49.730356&15.87&0.34731&10.4494&EW&E\\
95&395403867625020672&2.444296&51.015019&14.59&0.08088&8.5068&EW&RS\\
96&394620053267032320&3.777182&50.981245&17.69&0.58553&6.409&EW&EW\\
97&395376757790700032&3.306434&51.127022&14.64&0.04435&10.2743&EW&\\
98&394625134212185344&3.452258&51.102212&16.96&0.09544&6.3836&EW&EW$\vert$S\\
99&393778892515263744&1.920853&50.338218&16.39&0.14084&6.8291&EW&EW\\
100&394596791724300800&4.115014&50.732622&15.73&0.16436&8.0261&EW&EW\\
	\hline
\end{tabular}
\end{table*}

\begin{table*}
\begin{tabular}{c|c|c|c|c|c|c|c|c|c}
\hline
№&GAIA DR3 source id&RA&Dec&mag G&Amp&Per (days)&Type&Type-0\\	
\hline
101&394625306010871296&3.530354&51.098191&17.93&0.54707&6.6844&EW&EW\\
102&395365414774910848&2.749469&51.141326&17.84&0.15113&8.4312&EW&RS$\vert$ECL\\
103&393768167977177216&2.051434&50.02236&19.11&0.25899&7.0498&EW&EW\\
104&393752980972608768&2.234195&49.833612&16.49&0.0499&20.9756&EW&RS\\
105&394608950775225216&4.009876&50.955858&17.12&0.19045&5.4978&EW&EW\\
106&395407715915683072&2.230459&51.06545&15.9&0.15665&12.6474&EW&E\\
107&394984880672072576&4.07643&51.057003&17.3&0.32304&8.4036&EW&EW\\
108&2262463706381878016&285.023464&70.339249&14.98&0.10411&6.0866&EW&EW\\
109&2265362053392441344&282.858106&70.260736&14.78&0.36585&8.624&EW&EW\\
		\multicolumn{8}{c}{\textbf{Class: Pulsating (short-period variables)}}\\
110&395230557103838208&3.135785&50.225591&13.58&0.04539&2.79&DGS&\\
111&395228426800132608&3.370289&50.236349&13.8&0.03664&2.4&DGS&DGS\\
112&395225231341646336&2.586567&50.368908&14.16&0.02182&1.2&DGS&DGS\\
113&395216637105570048&2.53896&50.215033&13.64&0.02755&1.7&DGS&DGS\\
114&394475639286763776&3.460441&50.100083&17.05&0.22323&3.644&DGS&DGS\\
115&393709314043417472&2.70691&49.874564&13.65&0.02332&2.9738&DGS&DGS\\
116&395289621492439936&1.937173&50.621697&14.03&0.02181&0.85&DGS&DGS\\
117&394602048764259840&4.172585&50.762445&14.36&0.03554&1.9&DGS&DGS\\
118&2262857606422311040&285.228549&70.782716&14.11&0.0444&1.1234&DGS&DGS\\
119&2265495266098115200&283.000759&70.881247&12.41&0.0298&1.2&DGS&DGS\\
120&2262444052611330432&283.972003&70.085719&13.59&0.02804&1.2&DGS&DGS\\
121&395244159258089472&2.868577&50.372147&17.89&1.01857&15.0053&RR&RR\\
122&2262481401647182208&284.258035&70.542965&15.96&1.45256&11.7579&RR&RR\\
			\multicolumn{8}{c}{\textbf{Class: Eclipsing, Pulsating, or Rotating}}\\
123&395234439752169344&3.05832&50.420052&14.25&0.00718&8.0188&rot&\\
124&395233031002910848&3.074812&50.355225&15.54&0.00684&16.254&EW$\vert$rot$\vert$DGS$\vert$RR&DGS\\
125&395241513558355968&3.145331&50.510285&18.58&0.11171&14.02&EW$\vert$rot$\vert$DGS$\vert$RR&\\
126&395247148555693952&2.826295&50.482282&18.72&0.10756&31.7477&EW$\vert$rot$\vert$DGS$\vert$RR&\\
127&395247285994643456&2.784514&50.477077&17.78&0.06658&5.703&EW$\vert$rot$\vert$DGS$\vert$RR&EW\\
128&395247698311551744&2.828034&50.535027&18.73&0.13324&11.54&EW$\vert$rot$\vert$DGS$\vert$RR&EW\\
129&395229487649686144&3.257887&50.272096&18.82&0.10328&8.6011&EW$\vert$rot$\vert$DGS$\vert$RR&RS\\
130&395219248447600640&2.743739&50.244472&17.76&0.04376&15.9492&EW$\vert$rot$\vert$DGS$\vert$RR&\\
131&395222169025392000&2.665364&50.269477&19.1&0.19956&8.3&EW$\vert$rot$\vert$DGS$\vert$RR&EW\\
132&395341225518297600&3.346583&50.765682&18.34&0.07729&7.5188&EW$\vert$rot$\vert$DGS$\vert$RR&EW$\vert$DGS\\
133&394481613585244928&3.720843&50.307102&17.01&0.07133&7.3288&EW$\vert$rot$\vert$DGS$\vert$RR&RS$\vert$DGS\\
134&395265363517361024&2.367273&50.322496&17.14&0.15709&36.755&EW$\vert$rot$\vert$DGS$\vert$RR&SL\\
135&395278347196623488&2.198794&50.603909&18.42&0.17723&5.79&EW$\vert$rot$\vert$DGS$\vert$RR&E\\
136&394599196905954432&3.882717&50.732298&14.15&0.02923&27.5746&EW$\vert$rot$\vert$DGS$\vert$RR&DGS\\
137&395289788989565440&1.929221&50.659614&18.31&0.11733&7.5378&EW$\vert$rot$\vert$DGS$\vert$RR&EW\\
138&394984743233106048&4.069788&51.040912&17.98&0.10494&7.2828&EW$\vert$rot$\vert$DGS$\vert$RR&EW\\
139&394429769036291840&4.011307&49.742144&13.54&0.18438&13.1884&EW$\vert$rot$\vert$DGS$\vert$RR&EW$\vert$RR\\
140&2262865337363426688&284.555622&70.751203&14.37&0.03469&27.419&EW$\vert$rot$\vert$DGS$\vert$RR&n/a\\
141&2262491773992539008&283.966466&70.67798&18.13&0.13379&9.874&EW$\vert$rot$\vert$DGS$\vert$RR&EW\\
142&2262485486160392832&283.983567&70.610705&17.84&0.07613&19.87&EW$\vert$rot$\vert$DGS$\vert$RR&BY\\
143&2262828946104484608&286.792306&71.022917&16.84&0.10555&5.534&EW$\vert$rot$\vert$DGS$\vert$RR&EW\\
144&2262444739806093696&283.857291&70.092501&17.02&0.06864&6.516&EW$\vert$rot$\vert$DGS$\vert$RR&EW\\
145&2262430136917524352&284.348247&69.940137&19.99&0.23023&5.76&EW$\vert$rot$\vert$DGS$\vert$RR&\\
			\multicolumn{8}{c}{\textbf{Class: Pulsating (long-period variables)}}\\
146&394576622557648128&3.61672&50.394126&14.57&0.18479&771.36&LPV&RS\\
147&394475845441467648&3.441362&50.133345&19.4&0.08073&79.0&LPV&\\
148&395350644388701312&2.792577&50.775497&16.24&0.05701&330.72&LPV&SL\\
149&2262855201241187968&284.889344&70.810036&16.7&0.0853&403.5&LPV&BY$\vert$RS\\
150&2262438898650595840&284.492419&70.112149&13.78&0.1333&150.6&LPV&BY\\
	\hline
	\end{tabular}
\end{table*}

\bibliographystyle{apalike}
\small{
\begin{spacing}{0.5}
\bibliography{Vars}
\end{spacing}
}

\end{document}